\documentclass [letter,12pt, twoside] {report}
\usepackage[dvips]{graphicx}
\usepackage{amsfonts,amsmath,amssymb,amscd,array,euscript}

\makeatletter

\renewcommand{\section}{\@startsection{section}{1}%
{\parindent}{3.5ex plus 1ex minus .2 ex} {1.5 ex plus .2 ex}{\large\bf}}

\renewcommand{\subsection}{\@startsection{subsection}{1}%
{\parindent}{3.5ex plus 1ex minus .2 ex} {.5 ex plus .2 ex}{\normalsize\bf}}

\renewcommand{\sectionmark}[1]{}

\newcommand{\l@abcd}[2]{#1\dotfill #2 \\}

\makeatother

\newcommand {\ld} {\ldots}

\newcommand {\bge} {\begin{equation}}
\newcommand {\ee} {\end{equation}}
\newcommand {\bgen} {\begin{equation*}}
\newcommand {\een} {\end{equation*}}
\newcommand {\bgml} {\begin{multline}}
\newcommand {\eml} {\end{multline}}
\newcommand {\bgmln} {\begin{multline*}}
\newcommand {\emln} {\end{multline*}}   
\newcommand {\bgg} {\begin{gather}}
\newcommand {\eg} {\end{gather}}     
\newcommand {\bga} {\begin{array}}
\newcommand {\ea} {\end{array}}
\newcommand {\bgp} {\begin{picture}}
\newcommand {\ep} {\end{picture}}
\newcommand {\bgc} {\begin{center}}
\newcommand {\ec} {\end{center}}
\newcommand {\bgt} {\begin{tabular}}
\newcommand {\et} {\end{tabular}}

\newcommand {\nin}{\noindent}

\newcommand {\sms} {\smallskip}
\newcommand {\ses} {\medskip}
\newcommand {\mes} {\medskip}

\newcommand{\bea}{\begin{eqnarray}}
\newcommand{\eea}{\end{eqnarray}}

\renewcommand{\theequation}
{\arabic{equation}}

\def\2#1#2#3{{#1}_{#2}\hspace{0pt}^{#3}}
\def\3#1#2#3#4{{#1}_{#2}\hspace{0pt}^{#3}\hspace{0pt}_{#4}}
\newcounter{sctn}
\def\sec#1.#2\par{\setcounter{sctn}{#1}\setcounter{equation}{0}
                  \noindent{\bf\boldmath#1.#2}\bigskip\par}

\newcommand{\bib}{\bibitem}

\def\thebibliography#1{\subsubsection*{References}
\list
  {[\arabic{enumi}]}{\settowidth\labelwidth{[#1]}\leftmargin\labelwidth
  \advance\leftmargin\labelsep
  \usecounter{enumi}}
  \def\newblock{\hskip .11em plus 0.33em minus -.07em}
  \sloppy
  \sfcode`\.=1000\relax}
 
 \catcode `\@=12
\newcommand {\etit}[1] {\bgc{\textbf{\large{#1}}} \mes\mes\ec}  
\newcommand {\eauth}[2] {\bgc{\textbf{\large{#1}}\\ \mes\mes \emph{#2}} \mes\mes\ec}  
\newcommand {\esect}[1] {\mes\mes\mes\mes\nin{\textbf{#1}} \mes\mes}  

\begin {document}

\newcounter{\theequation}{\setcounter{equation}{0}}

 \etit{QUATERNIONS AND BIQUATERNIONS: ALGEBRA, GEOMETRY \\ \mes
 AND PHYSICAL THEORIES}

\eauth{A. P. Yefremov} {Russian University of people friendship \\
a.yefremov@rudn.ru}

{\small The review of modern study of algebraic, geometric and differential
properties of quaternionic (Q) numbers with their applications. Traditional and
"tensor"\, formulation of Q-units with their possible representations are
discussed and groups of Q-units transformations leaving Q-multiplication rule
form-invariant are determined. A series of mathematical and physical
applications is offered, among them use of Q-triads as a moveable frame,
analysis of Q-spaces families, Q-formulation of Newtonian mechanics in
arbitrary rotating frames, and realization of a Q-Relativity model comprising
all effects of Special Relativity and admitting description of kinematics of
non-inertial motion. A list of "Quaternionic Coincidences"\, is presented
revealing surprising interconnection between basic relations of some physical
theories and Q-numbers mathematics.}

\esect{Introduction}

The discovery of quaternionic (Q) numbers dated by 1843 is usually attributed
to Hamilton \cite{ham1, ham2}, but in the previous century Euler and Gauss made
a contribution to mathematics of Q-type objects; moreover Rodriguez offered
multiplication rule for elements of similar algebra \cite{stroik, burbaki,
bogolub}. Active opposition of Gibbs and Heaviside to Hamilton's disciples gave
a start to the modern vector algebra, and later to vector analysis, and
quaternions practically ceased to be a tool of mathematical physics, despite of
exclusive nature of their algebra confirmed by Frobenius theorem. At the
beginning of 20 century last bastion of Q-numbers amateurs, "Association for
the Promotion of the Study of Quaternions", was ruined.  The only reminiscence
of once famous hypercomplex numbers was the set of Pauli matrices. Later on
quaternions appeared incidentally as a mathematical mean for alternative
description of already known physical models \cite{klien, rastall} or due to
surprising simplicity and beauty they were used to solve rigid body cinematic
problems \cite{branets}. An interest to quaternionic numbers essentially
increased in last two decades when a new generation of theoreticians started
feeling in quaternions deep potential yet undiscovered (e. g.
\cite{horwitz}\,--\,\cite{bisht}).

This work is an attempt to give more systematic overview of contemporary state
of Q-number mathematics, its applications to physical theories and possible
perspectives in this area. In the context some quite specific even surprising
physical models, but worth to pay attention to, are shortly discussed.

The review arranged as follows. In section 1 general relations of the
quaternionic algebra are briefly described in the traditional hamiltonian
formulation as well as in tensor-like format. Section 2 is devoted to
description of structure of three "imaginary"\, quaternionic units. In section
3 the elements of differential Q-geometry are given with examples of their
mathematical application. Section 4 comprises Q-formulation of Newtonian
mechanics in the rotating frames of reference. Quaternionic Relativity Theory
with a number of cinematic relativistic effects is found in section 5. Section
6 contains the list of "Great Quaternion Coincidences"\, and final discussion.


 \esect{1. Algebra of quaternions}

 {\it Traditional approach}
 \ses

According to Hamilton, a quaternion is a mathematical object of
the form
 $$Q \equiv a + b{\bf i} + c{\bf j} + d{\bf k},$$
where $a, b, c, d$ are real numbers, $a$ is a coefficient at real unit "1", and
${\bf i, j, k}$ -- three imaginary quaternion units. The multiplication rule
for these units given by Hamilton and often used in literature is
 \begin{gather*}
  1{\bf i} = {\bf i}1 \equiv {\bf i}, \qquad
 1{\bf j} = {\bf j}1 \equiv {\bf j}, \qquad
 1{\bf k} = {\bf k}1 \equiv {\bf k},\\
 {\bf i}^2  = {\bf j}^2  = {\bf k}^2  =  - 1,\\
 {\bf ij} =  - {\bf ji} = {\bf k}, \qquad
 {\bf jk} =  - {\bf kj} = {\bf i}, \qquad
 {\bf ki} =  - {\bf ik} = {\bf j}
 \end{gather*}
These very cumbersome equations mean, that Q-multiplication loses a
commutativity.
 $$Q_1 Q_2  \ne Q_2 Q_1,$$
so that a notion of the right and the left multiplication appears, but it
remains associative.
 $$(Q_1 Q_2 )Q_3 = Q_1 (Q_2 Q_3 ).$$
Two rather different algebraic parts are separated naturally in a quaternion,
these once could be denoted as scalar:
$$scal~Q = a,$$
and vector
 $$vect~Q = b{\bf i} + c{\bf j} + d{\bf k}.$$
Addition (subtraction) of quaternions is performed by components, scalar and
vector parts are added (subtracted) separately. With respect to addition the
Q-algebra is commutative and associative.

Further step is quaternion conjugation introduced similarly to that of the
complex numbers
 $$\bar{Q} \equiv scal~Q - vect~Q = a - b{\bf i} - c{\bf j} - d{\bf k},$$
modulus of a Q-number is defined as
 $$\left| Q \right| \equiv \sqrt {Q\bar Q}  = \sqrt {a^2  + b^2  + c^2  + d^2 }.$$
This permit to formulate a quaternionic division being as multiplication
"right"\, and "left"
 $$ Q_L  = \frac{{Q_1 \bar Q_2 }}{{\left| {Q_2 } \right|^2 }},  \qquad
 Q_R  = \frac{{\bar Q_2 Q_1 }}{{\left| {Q_2 } \right|^2 }}. $$
Definition of Q-modulus enhances the famous four squares identity
 $$\left| {Q_1 Q_2 } \right|^2  = \left| {Q_1 } \right|^2 \left| {Q_2 } \right|^2.$$
Due to the properties mentioned above the Q-numbers form the algebra, which
belongs to the elite group of four the so-called exclusive -- "very good"\, --
algebras: of real, complex, quaternionic numbers and the octonions (Frobenious
and Horwits theorems of 1878-1898 \cite{kantor}).

Special attention should be paid to Q-units representations. In terms of
Hamilton real unit is simply 1 while three imaginary units similarly to complex
numbers algebra are denoted as {\bf i, j, k}. Later a simple $2\times 2$
matrices representation of these units was revealed
 $${\bf i} =  - i\left( {\begin{array}{*{20}c}
   0 & 1  \\
   1 & 0  \\
\end{array}}\right), \qquad
 {\bf j} =  - i\left( {\begin{array}{*{20}c}
   0 & { - i}  \\
   i & 0  \\
\end{array}}\right), \qquad
 {\bf k} =  - i\left( {\begin{array}{*{20}c}
   1 & 0  \\
   0 & { - 1}  \\
\end{array}}\right). $$
This representation of course is not unique. Here is a simple example. If in
the above expressions imaginary unit $i$ of complex numbers is represented as
$2\times 2$ with real elements
 $$i = \left({\begin{array}{*{20}c}
   0 & 1  \\
   { - 1} & 0  \\
 \end{array}} \right), $$
then three vector Q-units turn out to be represented by real $4\times 4$
matrices. The procedure of the matrix rank duplication can obviously be
continued further.

 \ses\ses
 {\it "Tensor"\, form and representations}
 \ses

If each Q-unit is endowed with its proper number (as components of a tensor)
 $$({\bf i, j, k}) \to ({\bf q}_1 ,{\bf q}_2 ,{\bf q}_3 ) = {\bf q},
 \qquad k, j, k, l, m, n, \ld = 1, 2, 3,$$
then quaternionic multiplication rule acquires compact form
 $$1{\bf q}_k  = {\bf q}_k 1 = {\bf q}_k, \qquad
 {\bf q}_j {\bf q}_k  =  - \delta _{jk}  + \varepsilon _{jkn} {\bf q}_n,$$
where $\delta_{kn}$ and $\varepsilon_{knj}$ -- respectively,
3-dimension (3D) symbols Kronecker and Levi-Chivita.

It is easy to show that a number of the Q-units representations even only by $2
\times 2$ matrices is infinite. Indeed for any $2 \times 2$ matrices with
properties
 $$A = \left( {\begin{array}{*{20}c}
   a & b  \\
   c & { - a}  \\
 \end{array}}\right), \qquad
 B = \left( {\begin{array}{*{20}c}
   d & e  \\
   f & { - d}  \\
 \end{array}}\right), \qquad
 Tr A = Tr B = 0,$$
the first two Q-units can be constructed as follows
 $${\bf q}_1  = \frac{A}{{\sqrt {\det A} }}, \qquad
 {\bf q}_2  = \frac{B}{{\sqrt {\det B} }}, $$
while the third one is
 $${\bf q}_3  \equiv {\bf q}_1 {\bf q}_2  = \frac{{AB}}{{\sqrt {\det A\det B}
 }} \qquad \mbox{provided that} ~~ Tr(AB) = 0.$$
The scalar unit is always invariant:
 $$1 = \left( {\begin{array}{*{20}c}
   1 & 0  \\
   0 & 1  \\
\end{array}} \right).$$

 \clearpage
{\it Transformations of Q-units and invariancy of the
multiplication rule}

 \ses
{\it a. Spinor-type transformations}

 \ses
If $U$ is an operator changing at once all the units, and there is an inverse
operator $U^{-1}:~~ UU^{-1}=E$, then transformations
 $${\bf q}_{k'}  \equiv U{\bf q}_k U^{ - 1} \quad \mbox{and} \quad
1' \equiv U1U^{ - 1}  = E1 = 1$$ retain the multiplication rule
 $$1{\bf q}_k  = {\bf q}_k 1 = {\bf q}_k, ~~
 {\bf q}_j {\bf q}_k  =  - \delta _{jk}  + \varepsilon _{jkn} {\bf q}_n$$
form-invariant
 $${\bf q}_{k'} {\bf q}_{n'}  = U{\bf q}_k U^{ - 1} U{\bf q}_n U =
 U\delta _{kn} U^{ - 1}  + \varepsilon _{knj} U{\bf q}_j U^{ - 1} =
 \delta _{kn}  + \varepsilon _{knj} {\bf q}_{j'}. $$
Such operator can be represented for example by  $2\times 2$ matrix
 $$ U = \left( {\begin{array}{*{20}c}
   a & b  \\
   c & d  \\
 \end{array}} \right), \quad \det U = 1, $$
or unimodular quaternion,
 $$U = \frac{{a + d}}{2} + \sqrt {1 - \left( {\frac{{a + d}}{2}} \right)^2 } {\bf q},$$
where
 $${\bf q} \equiv \left( {\sqrt {1 - \left( {\frac{{a + d}}{2}} \right)^2 } } \right)^{ - 1}
 \left( {\begin{array}{*{20}c}
   {\frac{{a - d}}{2}} & b  \\
   c & { - \frac{{a - d}}{2}}  \\
 \end{array}} \right).$$
In general this transformation contains 3 independent complex parameter (or 6
real ones), then $U \in SL(2,C)$. In special case of only three real
parameters, then $U \in SU(2)$.

 \mes
{\it b. Vector type transformations}

Vector Q-units can be transformed by $3\times 3$ matrix $O_{k'n}$
 $${\bf q}_{k'}  = O_{k'n} {\bf q}_n.$$
The requirement of Q-multiplication form-invariance forces the transformation
matrix to be orthogonal and unimodular
 $$O_{k'n} O_{j'n}  = \delta _{kn} \Rightarrow
 O_{nk'}^{ - 1}  = O_{k'n}, ~~ \det O = 1. $$
This transformation in general has 6 independent real parameters, then $O \in
SO(3,C)$. In the special case of three parameters $O \in SO(3,R)$. Below a
variant of representation of the transformation matrix $O$ is given with
$x,y,z$ being arbitrary real or complex functions
 $$ O = \left( {\begin{array}{*{20}c}
   {\sqrt {1 - x^2  - z^2 } }
   & { - \frac{{x\sqrt {1 - y^2  - z^2 }  + yz\sqrt {1 - x^2  - z^2 } }}{{1 - z^2 }}}
   & {\frac{{xy - z\sqrt {1 - x^2  - z^2 } \sqrt {1 - y^2  - z^2 } }}{{1 - z^2 }}}  \\
   x & {\frac{{\sqrt {1 - x^2  - z^2 } \sqrt {1 - y^2  - z^2 }  - xyz}}{{1 - z^2 }}}
   & {\frac{{ - y\sqrt {1 - x^2  - z^2 }  - xz\sqrt {1 - y^2  - z^2 } }}{{1 - z^2 }}}  \\
   z & y & {\sqrt {1 - y^2  - z^2 } }  \\
 \end{array}} \right). $$
This matrix can be represented as a product of three irreducible
multipliers
 $$O = \left( {\begin{array}{*{20}c}
   {\sqrt {\frac{{1 - x^2  - z^2 }}{{1 - z^2 }}} } & { - \frac{x}{{\sqrt {1 - z^2 } }}} & 0  \\
   {\frac{x}{{\sqrt {1 - z^2 } }}} & {\sqrt {\frac{{1 - x^2  - z^2 }}{{1 - z^2 }}} } & 0  \\
   0 & 0 & 1  \\
 \end{array}} \right)\left( {\begin{array}{*{20}c}
   {\sqrt {1 - z^2 } } & 0 & { - z}  \\
   0 & 1 & 0  \\
   z & 0 & {\sqrt {1 - z^2 } }  \\
 \end{array}} \right)\left( {\begin{array}{*{20}c}
   1 & 0 & 0  \\
   0 & {\sqrt {\frac{{1 - y^2  - z^2 }}{{1 - z^2 }}} } & { - \frac{y}{{\sqrt {1 - z^2 } }}}  \\
   0 & {\frac{y}{{\sqrt {1 - z^2 } }}} & {\sqrt {\frac{{1 - y^2  - z^2 }}{{1 - z^2 }}} }  \\
 \end{array}} \right).$$
after substitutions
 $z \equiv \sin {\rm B}, ~
 x \equiv  - \sin {\rm A}\cos {\rm B}, ~
 y \equiv  - \sin \Gamma \cos {\rm B}, $
where $A, B, \Gamma$ -- are complex "angles", it takes the form
 $$O = \left( {\begin{array}{*{20}c}
   {\cos {\rm A}} & {\sin {\rm A}} & 0  \\
   { - \sin {\rm A}} & {\cos {\rm A}} & 0  \\
   0 & 0 & 1  \\
 \end{array}} \right)\left( {\begin{array}{*{20}c}
   {\cos {\rm B}} & 0 & { - \sin {\rm B}}  \\
   0 & 1 & 0  \\
   {\sin {\rm B}} & 0 & {\cos {\rm B}}  \\
 \end{array}} \right)\left( {\begin{array}{*{20}c}
   1 & 0 & 0  \\
   0 & {\cos \Gamma } & {\sin \Gamma }  \\
   0 & { - \sin \Gamma } & {{\rm cos}\Gamma }  \\
 \end{array}} \right) = O_3^{\rm A} O_2^{\rm B} O_1^\Gamma.$$
If the angles are real: ${\rm A} = \alpha, ~ {\rm B} = \beta, ~ \Gamma  =
\gamma,$ then this transformation is an ordinary vector rotation consisting of
three simple rotations around numbered orthogonal axes: $O \Rightarrow R, R =
R_3^\alpha R_2^\beta R_1^\gamma.$ Correlation between related "spinor"\, and
"vector"\, transformations is easily determined:
 $$ O_{k'n}  =  - \frac{1}{2}Tr(U{\bf q}_k U^{ - 1} {\bf q}_n ), \qquad
 U = \frac{{1 - O_{k'n} {\bf q}_k {\bf q}_n }}{{2\sqrt {1 + O_{mm'} } }}. $$

  \ses

{\it Q-geometry in three dimensional space}

 \ses

Hamilton was the first to note that triad of Q-units behaves as three strictly
tied unit vectors (with length $i$) initiating Cartesian coordinate system,
somewhat exotic because of its "imaginarity". Due to the fact the Q-triad in
3D-space $({\bf q}_1 ,{\bf q}_2 ,{\bf q}_3 )$ will be called 'quaternionic
basis' (Q-basis). Now Q-units transformations have apparent geometrical sense
of various rotations of the Q-basis. An example: a simple rotation by real
angle $\alpha$ around axis \# 3
 $$ {\bf q'} = R_3^\alpha  {\bf q}. $$
Notion of Q-basis helps to introduce 3D quaternionic vectors (Q-vectors),
defined as
 $$ {\bf a} = a_k {\bf q}_k, $$
here all its components $ a_{k}$ are real. The most important property of
Q-vector -- is its invariancy with respect to vector transformations from the
group SO(3,R)
 $$ {\bf a'} = a_{k'}{\bf q}_{k'}  =
 a_{k'} R_{k'j} {\bf q}_j  = a_j {\bf q}_j  = {\bf a}.$$
The projection of Q-vector onto arbitrary coordinate axis (represented by any
different Q-unit) can be found in two ways. First, if at least one set of
projections of Q-vector and rotation matrices $R_{nk'}$ are known then
projections of this vector on rotated axis are immediately found
 $$ a_{k'}  = a_n R_{nk'}. $$
The second approach is related to existence of internal structure of the
Q-units; a brief analysis of it is given in the next section.

 \clearpage

 \esect{2. Structure of quaternionic "imaginary"\, units}

{\it Eigenfunctions of Q-units} \cite{yefr1}

 \ses

Each vector Q-unit can be thought of as operator, so eigenfunctions and
eigenvalues problem can be formulated for it
 $${\bf q}\psi  = \lambda \psi, \qquad
 \varphi {\bf q} = \mu \varphi.$$
The solution of this problem are the eigenvalues ("imaginary length"\, of
Q-unit with division by parity)
 $$\lambda  = \mu  =  \pm i,$$
and two sets of eigenfunctions(one for each parity), possible given by columns
$\psi^\pm$ and rows $\varphi^\pm$, being the functions of components ${\bf q}$.

Here is an example explicit form of eigenfunction: for the Q-unit represented
by matrix
 $$ {\bf q}=  - \frac{i}{T}\left( {\begin{array}{*{20}c}
   a & b  \\
   c & { - a}  \\
 \end{array}} \right),$$ \\
where $T \equiv a^2  + bc \ne 0, b \ne 0, c \ne 0$, its eigenfunctions are
defined as
 $$\varphi^\pm = x\left( {\begin{array}{*{20}c}
   1 & { \pm \frac{b}{{T \pm a}}}  \\
 \end{array}} \right), \qquad
 \psi ^ \pm   = y\left( {\begin{array}{*{20}c}
   1  \\
   { \mp \frac{c}{{T \pm a}}}  \\
 \end{array}} \right), $$
where $x, y$ are arbitrary complex factors.

The freedom of components, arising in the calculations is reduced by convenient
normalization condition
 $$\varphi ^ \pm  \psi ^ \pm   = 1,$$
while the eigenfunctions orthogonality (by parity) is an inherited property
 $$\varphi^\mp \psi^\pm = 0.$$
One can construct tensor products of eigenfunctions and obtain $2\times2$
matrices
 $$ C^\pm \equiv \psi^\pm \varphi^\pm, $$
possessing a properties reciprocal with respect to the ones of vector
$\textbf{q}$:
 $$ \det C = 0, ~~ Tr\,C = 1,$$
whereas
 $$\det {\bf q} = 1, ~~ Tr\,{\bf q} = 0.$$
Matrix C is idempotent
 $$C^n  = C,$$
and can be expressed throw their own unit Q-vector
 $$C^ \pm   = \frac{{1 \pm i{\bf q}}}{2}.$$
When inversed the latter expression gives information about internal structure
of Q-unit
 $${\bf q} =  \pm i(2C^ \pm - 1) = \pm i(2\psi^\pm \varphi^\pm - 1),$$
which turns out to consist of a combination of its eigenfunctions and scalar
units.

Since each Q-unit has its own eigenfunctions the Q-triad as a whole possesses
unique set of eigenfunctions $\{\varphi_{(k)}^\pm,\psi_{(k)}^\pm\}$. There is
an interesting algebraic observation concerning this set. Three Q-units are
interrelated by obviously nonlinear combination -- multiplication e. g.
 $$ {\bf q}_3 = {\bf q}_1 {\bf q}_2, $$
but it is easy to show that corresponding eigenfunctions depend on each other
linearly:
 $$ \varphi_{(3)}^\pm = \sqrt {\mp i} \varphi_{(1)}^\pm \pm \sqrt i \varphi_{(2)}^ \pm, ~~
 \psi _{(3)}^ \pm   = \sqrt { \pm i} \psi _{(1)}^ \pm   \pm \sqrt { - i} \psi _{(2)}^ \pm.$$
Q-eigenfunctions help to represent a spinor-type transformation of Q-units
retaining Q-multiplication invariant in the familiar form
 $$ \psi _{(k')}^ \pm   = U\psi _{(k)}^ \pm, ~~
 \varphi _{(k')}^ \pm   = \varphi _{(k)}^ \pm  U^{ - 1}, $$
so that the eigenfunctions can be regarded as a set of specific spinor
functions, allowing in subject in general to $SL(2C)$ transformations. Yet
another mathematical observation should be noted: from pairs of eigenfunctions,
belonging to different Q-units of one triad and having one parity, one can
construct 24 scalar invariants $SL(2C)$ group; these invariants are real or
complex numbers, e. g.:
 $$ \sigma_{12}^+ \equiv \varphi_{(1)}^+ \psi_{(2)}^+ =
 \sqrt{- \frac{i}{2}} = \frac{{1 - i}}{2}. $$

 \ses\ses
 {\it Quaternionic eigenfunctions as projectors}

 \ses

Eigenfunctions act on their own Q-basis as following
 $$ \varphi_{(1)}^\pm {\bf q}_1 \psi_{(1)}^\pm = \pm i, ~~
 \varphi_{(1)}^\pm {\bf q}_2 \psi_{(1)}^\pm = 0, ~~
 \varphi_{(1)}^\pm {\bf q}_3 \psi_{(1)}^\pm = 0, $$
or in general
 $$ \varphi_{(k)}^\pm {\bf q}_n \psi_{(k)}^\pm = \pm i\delta_{kn}
 \qquad \mbox{(no summation by k)}. $$

It looks like that eigenfunctions select a projection of the unit Q-vector,
generating them. This idea is confirmed by an example of an action of
eigenfunctions of one Q-basis onto the vectors of the rotated Q-basis
 $$ \varphi_{(k)}^\pm {\bf q}_{n'} \psi_{(k)}^\pm
 = \varphi_{(k)}^\pm R_{n'm}{\bf q}_m \psi_{(k)}^\pm = \pm iR_{n'k}
 = \pm i\cos \angle ({\bf q}_{n'} ,{\bf q}_k ) \quad
 \mbox{(no summation by k)}, $$
the result of the action is 'nearly' projection of Q-basis ${{\bf q'}}$ on
${\bf q}$. It is convenient to denote precise projection as
 $$ \left\langle {{\bf q}_{{\bf n'}}} \right\rangle_k \equiv
 \mp i\varphi_{(k)}^\pm {\bf q}_{n'} \psi_{(k)}^ \pm
 = \cos \angle({\bf q}_{n'},{\bf q}_k) \qquad
 \mbox{(no summation by k)}. $$

It is now easy to formulate rule of calculation of projection of a Q-vector a
onto arbitrary direction, defined by vector ${\bf q}_j$ (e. g. with help of
eigenfunctions of positive parity)
 $$\left\langle {\bf a} \right\rangle_j^ +
 \equiv - ia_{k'} \varphi_{(j)}^ + {\bf q}_{k'} \psi_{(j)}^ +
 = a_{k'} R_{k'j} = a_j \qquad
 \mbox{(no summation by j)}.$$
Thus quaternionic eigenfunctions with their own interesting properties are more
fundamental mathematical objects then Q-units and too can serve as useful tool
for practical purposes such as computing projections of Q-vectors.

 \clearpage
\esect{4. Differential Q-geometry}

{\it Quaternionic connection}

 \ses

If vectors of Q-basis are smooth functions of parameters
 ${\bf q}_k (\Phi_\xi)$ (index $\xi$ enumerates parameters), then
 $$d{\bf q}_k (\Phi)= \omega_{\xi~kj} {\bf q}_j d\Phi_\xi,$$
where an object $\omega_{\xi~kj}$ is called quaternionic connection.
Q-connection is antisymmetric in vector indices
 $$\omega_{\xi~kj}+ \omega_{\xi~jk}= 0,$$
and has the following number of independent components
 $$N = Gp(p - 1)/2,$$
where G is an number of parameters and p = 3 -- is a number of space
dimensions. If G = 6 [a case of group SO(3,C)], then N = 18; if G = 3 [a case
of group SO(3,R)], then N = 9. Q-connection can be calculated at least in three
ways:
 $$ \mbox{using vectors of Q-basis} \qquad \omega_{\xi~kn} = \left\langle
 {\frac{{\partial {\bf q}_k }} {{\partial \Phi _\xi  }}}
 \right\rangle_n^ +, $$
using matrices U from the group SL(2C) (general case) and special
representation of constant Q-units
 ${\bf q}_{\tilde k} = - i\sigma_k$,
where $\sigma_k$ -- Pauli matrices
 $$ \omega_{\xi~kn} = \left\langle {U^{ - 1}
 \frac{{\partial U}}{{\partial \Phi_\xi}}{\bf q}_{\tilde k}
 - {\bf q}_{\tilde k} U\frac{{\partial U^{ - 1}}}{{\partial \Phi_\xi}}}
 \right\rangle_n^+, $$
and, finally, using matrices $O$ from $SO(3,C)$ (in a general
case)
 $$\omega_{\xi~kn} = \frac{{\partial O_{k\tilde j}}}{{\partial \Phi_\xi}}O_{n\tilde j}.$$
All the formulae of course provide same result.

From the point of view of vector transformations a Q-connection is not a
tensor. If ${\bf q}_k = O_{kp'}{\bf q}_{p'}$, then transformed components of
connection are expressed throw original ones with addition of inhomogeneous
term
 $$ \omega_{\xi~kj} = O_{kp'} O_{jn'} \omega_{\xi~p'n'}
 + O_{jp'} \frac{{\partial O_{kp'}}}{{\partial \Phi_\xi}}.$$
In 3D space Q-connectivity has clear geometrical and physical treatment as
moveable Q-basis with behavior of Cartan 3-frame. Parameters of its ordinary
rotations can depend on spatial coordinates $\Phi_\xi = \Phi_\xi (x_k ),$ then
$\partial_n {\bf q}_k = \Omega_{nkj} {\bf q}_j$, then components of slightly
modified Q-connection
 $$ \Omega _{nkj} \equiv \omega_{\xi~kj} \partial_n \Phi_\xi$$
have a sense of Ricci rotation coefficients. Parameters can also depend on the
length of line of motion of the Q-basis or on the observer's time. Then
 $\Phi_\xi = \Phi_\xi(t), \partial_t{\bf q}_k = \Omega_{kj} {\bf q}_j$,
and components of Q-connection
 $$ \Omega _{kj}  \equiv \omega _{\xi~kj} \partial _t \Phi _\xi$$
became generalized angular velocities of rotations of the frame.

The typical examples of Q-frames and Q-connection are

a) Frene frame. For the smooth curve $x_{\tilde k}(s)$ defined in constant
basis the Frene frame is represented by the triad ${\bf q}_k$, obeying the
equations
 $$ \frac{d}{{ds}}{\bf q}_1 = R_I (s){\bf q}_2, ~
 \frac{d}{{ds}}{\bf q}_2 = - R_I (s){\bf q}_1 + R_{II}(s){\bf q}_3,~
 \frac{d}{{ds}}{\bf q}_3 = - R_{II}(s){\bf q}_2, $$
where the first and the second curvatures are
 $$ R_I = \Omega_{12}, R_{II} = \Omega _{23}. $$

b) Twisted straight line. For a given straight line $x_{\tilde 1}= u,~
x_{\tilde 2} = x_{\tilde 3} = 0$, one can construct a Q-basis associated with
it so that one vector is tangent to the line. In this case Q-connection is not
zero and represented the only component describing torsion (or rather twist) of
the line about itself.
 $$ {\bf q}_1  =  - i\left( {\begin{array}{*{20}c}
   1 & 0  \\
   0 & { - 1}  \\
 \end{array}} \right), \qquad
 {\bf q}_2  =  - i\left( {\begin{array}{*{20}c}
   0 & { - ie^{ - i\gamma (u)} }  \\
   {ie^{i\gamma (u)} } & 0  \\
 \end{array}} \right), \qquad
 \Omega _{23}  = \frac{{d\gamma }}{{du}}, $$
here $\gamma (u)$ is the angle, which is an arbitrary but smooth function of
the line length.

 \ses\ses

 {\it Quaternionic spaces}

 \ses

Tangent Q-space \cite{yefr3}. It is known that on every N-dimensional
differentiable manifold $U_{N}$ with coordinates $\{y^{A}\}$ one can construct
a tangent space $T_{N}$ with coordinates $\{ X^{(A)} \}$ so that $dX^{(A)} =
g_B^{(A)} dy^{B}$, where $g_{B}^{(A)}$ -- Lame coefficients. By an extra
rotation one can construct a tangent Q-space $T(U,{\bf q})$, with coordinates
$\{ x_{k} \}$, k = 1,2,3, which associated with Q-frame vectors.
 $$ dx_{k} = h_{k(A)} dX^{(A)} = h_{k(A)} g_{B}^{(A)} dy^{B},$$
where $h_{k(A)}$ are in general non-square matrices normalized by projectors of
the basic space onto 3D one or vice versa.

Proper quaternionic space itself ${\bf U}_{3}$ is defined as 3D-space, locally
identical to own tangent space $T({\bf U}_{3} ,{\bf q})$. The Q-space has the
following basic features. Its Q-metric represented by vector part of the
Q-multiplication rule ${\bf q}_{j} {\bf q}_{k} = - \delta_{jk} +
\varepsilon_{jkn} {\bf q}_{n}$ is nonsymmetric, its antisymmetric part is
Q-operator (matrix), so that every point ${\bf U}_{3}$ has internal
quaternionic structure. Q-connection ${\bf U}_{3}$ can be: (i) proper (metric)
$\Omega_{nkj} \equiv \omega_{\xi kj} \partial_{n} \Phi_{\xi}$, for variable
Q-basis it is always non zero, and (ii) affine (non-metric), independent from
Q-basis. Q-torsion does not vanish in both cases, whereas Q-curvature
 $r_{knab} = \partial_{a} \Omega_{bkn} - \partial_{b} \Omega_{akn}
 + \Omega_{ajn} \Omega_{bjk} - \Omega_{bjk} \Omega_{ajn}$
for the metric Q-connection identically equals zero, but can be present in the
space of affine Q-connection.

Once Q-space is introduced, there appears a new field of investigation of
differential manifolds and spaces. Thus in the preliminary classification of
Q-spaces based on presence and nature of curvature, torsion and non-metricity
at least 10 different families can be distinguish \cite{yefr3}. In addition
Q-spaces can be a nontrivial background for classical and quantum theories and
problems.

 \clearpage
 \esect{4. Newton mechanics in Q-basis}

 {\it Dynamics equations in rotating frame} \cite{yefr4}

\ses\ses

The Q-basis endowed with clock becomes a classical (non-relativistic) reference
system. For an inertial observer the dynamic equations of classical mechanics
can be written in constant Q-basis
 $$ m\frac{d^{2}}{dt^{2}}x_{\tilde{k}} {\bf{q}}_{\tilde{k}} =
 F_{\tilde{k}} {\bf{ q}}_{\tilde{k}}.$$
 $SO(3,R)$-invariance of two Q-vectors, the radius-vector
${\bf r} \equiv x_k {\bf q}_k$ and force ${\bf F}~\equiv~F_k {\bf q}_k$ allow
to represent these equations in Q-vector form
 $$ m\frac{{d^2 }}{{dt^2 }}(x_k {\bf q}_k ) = F_k {\bf q}_k, \qquad
 \mbox{or} \quad  m{\bf \ddot r} = {\bf F} $$
In explicit form these equations possess enough complicated
structure
 $$ m(\frac{{d^2 }}{{dt^2 }}x_n + 2\frac{d}{{dt}}x_k \Omega_{kn}
 + x_k \frac{d}{{dt}}\Omega_{kn} + x_k \Omega_{kj} \Omega_{jn}) = F_n$$
which nevertheless can be simplified and interpreted from physical points of
view. Due to antisymmetry of the connection (generalized angular velocity)
 $$ \Omega_j \equiv \Omega_{kn} \frac{1}{2}\varepsilon _{knj},
 \qquad \Omega_{kn} = \Omega_j \varepsilon_{knj}, $$
the dynamic equations can be rewritten in vector components
 $$ m(a_n + 2v_k \Omega_j \varepsilon_{knj} + x_k \frac{d}{{dt}}\Omega_j \varepsilon_{knj}
 + x_k \Omega_j \Omega_m \varepsilon_{jkp} \varepsilon_{mpn}) = F_n$$
or by conventional vector notation
 $$ m(\vec a + 2\vec \Omega \times \vec v + \dot \vec \Omega \times \vec r
 + \vec \Omega \times (\vec \Omega \times \vec r)) = \vec F.$$
Among left hand side terms one easily recognizes 4 classical accelerations:
linear, Coriolis, angular and centripetal. However this traditional
interpretation is good only for simple rotation; in the case of combination of
many Q-frame rotations number of components of generalized accelerations highly
increases, and the equations become much more complicated. However it is worth
noting that derivation of these equations for the most complicated rotations
with the help of Q-basis and Q-connection is extremely simple.

 \ses\ses
 {\it Samples of Q-formulation of problems of classical mechanics}
 \ses

'Chasing' Q-basis -- is a frame with one of its vectors, say  ${\bf q}_1$ is
always directed to observed particle. Dynamic equations for this case are
written in explicit form in following manner
 $$ \ddot r - r(\Omega_2^2 + \Omega_3^2 ) = F_1 /m,$$
 $$ 2\dot r\Omega_3 + r\dot \Omega_3 + r\Omega_2 \Omega_1 = F_2 /m,$$
 $$ 2\dot r\Omega_2 + r\dot \Omega_2 + r\Omega_1 \Omega_3 = - F_3 /m.$$
Components of Q-connection are defined as functions of angles of two rotations,
the first (an angle $\alpha$) -- around vector ${\bf q}_3$, the second (an
angle $\beta$)
 -- around ${\bf q}_2$
 $$ \Omega_1 = \dot \alpha \sin \beta, \qquad \Omega_2 = - \dot
 \beta, \qquad \Omega_3 = \dot \alpha \cos \beta.$$
The chasing Q-basis approach is convenient to solve a number of mechanical
problems related to rotations, some times very complicated, of observed objects
and systems of reference. Here is an illustration.

Rotating oscillator. One seeks for motion law $r(t)$ of a harmonic oscillator
(mass m, spring elasticity k) which has a freedom of motion along rigid smooth
rod rotating in the plane around one of its ends (here one end of the spring is
fixed) with angular velocity $\omega$; the equilibrium point is located at the
distance l from the rotation center, there is no gravity. Radial and tangent
dynamic equations in the chasing Q-basis ($F$ is unknown rod reaction force)
 $$ \ddot r - r\omega ^2 = - \frac{k}{m}(r - l), \qquad
 2\dot r\omega = \frac{1}{m}F, $$
admit the following family of solutions:
 $$ \mbox{(i)} \qquad r(t) = r_0 + v_0 t + at^2$$
mass moves away from the center of rotation with quadratic (or
linear) law,
 $$ \mbox{(ii)} \qquad r(t) = const + Ae^{iwt} + Be^{- iwt}, \qquad
 w \equiv \sqrt {k/m - \omega^2}$$
here are three different situations depending on a relation of the quantities
under the square root:

 -- r = const,

 -- harmonic oscillators,

 -- exponential motion away from the center of rotation.

It is interesting that the variants of rotating classical oscillator behavior
with $l=0$ are precisely similar to behavior of four known cosmological models
of Einstein-DeSitter-Friedman considered in the General Relativity.

 \esect{5. Construction of Quaternionic Relativity}

 {\it Hyperbolic rotations and biquaternions} \cite{yefr5}
 \ses

It was noted above, that SO(3,C)-transformations of Q-units admit pure
imaginary parameters. In this case rotations become hyperbolical (H -- from
hyperbolic); e. g. simple H-rotation ${\bf q'} = H_3^\psi {\bf q}$ is performed
by matrix of the form
 $$ H_3^\psi   = \left( {\begin{array}{*{20}c}
   {\cosh \psi} & { - i\sin \psi} & 0  \\
   {i\sin \psi} & {\cosh \psi} & 0  \\
   0 & 0 & 1  \\
 \end{array}} \right), $$
and $2 \times 2$ -matrices of Q-units representation are no longer
hermitian:
 $$ {\bf q}_{1'}  =  - i\left( {\begin{array}{*{20}c}
   0 & {e^\psi}  \\
   {e^{ - \psi}} & 0  \\
 \end{array}} \right). $$
This is the time to recall the notion of so called biquaternionic vectors (BQ).
BQ-vector is defined as Q-vector with complex components ${\bf u} = (a_k + ib_k
){\bf q}_k$. Obviously for vectors of this type the norm (or modulus) in
general can not be defined; but among all BQ-vectors there is a subset of
"good"\, elements with well definable norm by ${\bf u}^2 = b^2 - a^2$. These
vectors appear to be form-invariant with respect to transformations of subgroup
$SO(2,1) \subset SO(3,C)$, and in particular, with respect to simple
H-rotations ${\bf q'} = H{\bf q} {\bf u} = u_k {\bf q}_k = u_{k'} {\bf
q}_{k'}$, but only when reciprocally imaginary components
 $a_k b_k = 0$ are orthogonal to each other.

 \ses\ses
 {\it Quaternionic Relativity}
 \ses

The made above observation allows to suggest a space-time BQ-vector "interval"
 $$ d{\bf z} = (dx_k  + idt_k ){\bf q}_k, $$
with specific properties:

(i) Temporal interval is defined by imaginary vector,

(ii) space-time of the model appears to have six-dimensional (6D),

(iii) vector of the displacement of the particle and vector of corresponding
time change must always be normal to each other $dx_k dt_k = 0$.

In this case BQ-vector-interval is invariant under group $SO(2,1) \subset
SO(3,C)$, as well as of course its square (which differs from the square of
norm only by sign) $d{\bf z}^2 = dt^2  - dr^2$, the latter has precisely the
same form as a space-time interval of Special Relativity of Einstein. This
6D-model was initially named the Quaternionic Relativity. Temporal and spatial
variables symmetrically enter the expression of BQ-vector-interval, and the
Q-triad related to it describes relativistic system of reference $\Sigma \equiv
({\bf q}_1 ,{\bf q}_2 ,{\bf q}_3)$. Transition from one reference system to
another is performed with the help of 'rotational equations' of the type
$\Sigma ' = O\Sigma$ with matrix $O$ from the group $SO(2,1)$ is a product of
matrices of real and hyperbolic rotations. So the theory could also be named
(may be more correctly) 'Rotational Relativity'. The meaning of a simple
H-rotation is immediately revealed from the first line of equation $\Sigma'=
H_3^\psi \Sigma$ in the explicit form
 $$ i{\bf q}_{1'} = i\cosh \psi ({\bf q}_1 + \tanh \psi {\bf q}_2). $$
If like in Special Relativity $\cosh \psi = dt/dt'$, then
 $$ idt'{\bf q}_{1'} = idt({\bf q}_1 + V{\bf q}_2), $$
which describes motion of reference system ${\Sigma '}$ relative to $\Sigma$
with velocity $V$ along direction ${\bf q}_2$. It is easy to show that
$SO(2,1)$-rotations of Q-reference system enhance Lorenz coordinate
transformations and therefore all cinematic effects of Special Relativity.

It should be noted here that parameters of real and hyperbolic rotations can be
variable for instance dependent on observer's time. This hints to expect of the
discussed theory a possibility to describe non-inertial motions. Analysis of
the rotational equations confirms the expectation. Well-known relativistic
model of reference system constantly accelerated with respect to the inertial
one (hyperbolic motion), frequently found in literature and normally regarded
with use of assumption beyond frames of Special Relativity, in quaternionic
theory is solved naturally and fast not only from the inertial observer
viewpoint, but from position of accelerated frame too \cite{yefr6}.

The kinematic problem of other non-inertial motion -- relativistic circular
motion -- can be completely and precisely resolved by means of the rotation
equation $\Sigma ' = H_2^{\psi (t)} R_1^{\alpha (t)} \Sigma$, where ${\Sigma'}$
is reference system rotating along the circle around the immobile frame
${\Sigma}$. This problem also can be solved both from the point of view of
inertial observer, in this case the result has the form
 $$ t = \int {dt'\cosh \psi } (t'), \qquad
 \alpha(t) = \frac{1}{R}\int {dt'\tanh \psi (t')}, $$
 $$ a_{\tan}(t) = \frac{1}{{\cosh^2 \psi}}\frac{{d\psi}}{{dt}}, \qquad
 a_{norm}(t) = R\left({\frac{{d\alpha (t)}}{{dt}}} \right)^2, $$
and from the point of view of the observer in the reference system arbitrary
moving along circular orbit.

The solution of the problem of "classical"\, Thomas precession in the framework
of Special Relativity also needs additional assumptions, while in the
quaternionic theory has a single line form -- the first row of the matrix of
rotation equation $\Sigma''= R_1^{-\alpha(t)} H_2^\psi R_1^{\alpha(t)} \Sigma$,
in this case of course correct value of precession frequency is obtained
 $$ \omega _T  = (1 - \cosh \psi ) \approx  - \frac{1}{2}\omega V^2. $$

Moreover, the quaternionic theory of relativity appears to be able to describe
Thomas precession for the vectors moving along trajectories of general type.
The basic rotational equation in this case naturally generalized: $\Sigma'' =
R_{}^{- \theta(t)} H_{}^{\psi(t)} R_{}^{\theta (t)} \Sigma$, here $\theta (t)$
-- an angle of instant rotation. Requirement that an axis of hyperbolic
rotation be normal to the plane formed by the radius-vector of observed frame
and its velocity vector, is also significant. In this case formula of variable
frequency of general Thomas precession has the form
 $$\Omega _T  = \frac{d}{{dt}}(\theta  - \theta ').$$

An example of such Thomas precession is an apparent displacement of mercurial
perihelion, for which calculations give a value $\Delta \varepsilon  =
2,7''/100$ years.

Universal character of motion of the bodies (including non-inertial motions) in
the Quaternionic Relativity suggests seeking for new cinematic relativistic
effects. One is found in Solar System planets' satellites motion. Relative
velocity of the Earth and other planets changes with time and sometimes
achieves significant value comparable somehow to value of the fundamental
velocity. This can lead to discrepancy between calculated and observed from the
Earth cinematic magnitudes characterizing cyclic processes on this planet or
near it. In particular there must be a deviation of the planetary satellite
position. Such an angular difference is surprisingly found to be linearly
dependent upon the time of observation
 $$ \Delta \varphi  \approx \frac{{\omega V_E V_P }}{{c^2 }}t, $$
here $\omega$ is an angular velocity of satellite motion around the planet, V
-- are linear velocities of the Earth and the planet around the sun. The
magnitude of the effect is the following for the closest to the Jupiter and
"the fasters"\, Jupiter satellite $\Delta \varphi \cong 12'$ for 100
terrestrial years; for the Mars satellite (Phobos) $\Delta \varphi \cong 20'$
for 100 terrestrial years \cite{yefr7}. Both values are big enough for the
effect to be noticed in prolonged and precise observations.

One can say that space-time model and kinematics of the Quaternionic Relativity
are nowadays studied in enough details and can be used as an effective
mathematical tool for calculation of many relativistic effects. But respective
relativistic dynamic has not been yet formulated, there are no quaternionic
field theory; Q-gravitation, electromagnetism, weak and strong interactions are
still remote projects. However, there is a hope that it is only beginning of a
long way, and the theory will "mature". This hope is supported by observation
of number of remarkable "Quaternionic Coincidences"\, forming a discrete mosaic
of physical and mathematical facts; probably one day it will turn into a
logically consistent picture providing new instruments and extending our
insight of physical laws.

 \esect{6. Remarkable "quaternionic coincidences"}

There are, at least, five such coincidences (all of them given
below), noted by different authors in various time.

 \ses
 1. {\it The Maxwell equations as an conditions of the analyticity
  of functions of quaternionic variable.}
 \sms

In 1937 year Fueter \cite{fueter} noted, that Cauchy-Riemann $\partial
f/\partial z^* = 0$ equations defining the differentiability of complex
variable function and modeling physically a flat motion of liquid without
sources and whirls, have the following quaternionic analogue
 $$ \left({i\frac{\partial}{{\partial t}} - {\bf q}_{\tilde k}
 \frac{\partial}{{\partial x_{\tilde k} }}} \right){\bf H} = 0, \qquad
 {\bf H} = (B_{\tilde n}  + iE_{\tilde n} ){\bf q}_{\tilde n}. $$
Surprising fact is that the equations of classic Maxwell electrodynamics in
vacuum prove to be corresponding physical model
 $$ div\vec E = 0, \quad div\vec B = 0, \quad
 rot\vec E - \frac{{\partial \vec B}}{{dt}} = 0, \quad
 rot\vec B + \frac{{\partial \vec E}}{{dt}} = 0. $$

 \ses
 2. {\it Classical mechanics in the rotating reference systems.}
 \sms

The compact form of Newton equations in quaternion frame is described above in
section 4. Finally it should be stressed that the form of dynamics equations
naturally arising and externally primitive
  $$ m{\bf \ddot r} = {\bf F} $$
hides all possible combinations of rotations of reference systems or observed
bodies. Using differential quaternionic objects helps to easily obtain explicit
form of the equations whose elements have obvious physical meaning.

 \ses
 3. {\it The quaternionic theory of relativity.}
 \sms

1:1 isomorphism of the Lorenz group of Special Relativity and the group of
invariance of quaternionic multiplication $SO(3,C)$ leads to non-standard
theory of relativity with symmetric six-dimensional space-time. This theory
significantly differs from Einstein Special Relativity in origin, model,
possibilities and mathematical tools, but predicts absolutely similar cinematic
effects. Invariance of specific biquaternionic vector "interval"\, $d{\bf z} =
(dx_kn +i\,dt_k){\bf q}_k$  under subgroup $SO(2,1)$ with in general variable
parameters admits calculation of relativistic effects for non-inertial motion
of reference systems.

 \ses
 4. {\it Pauli equations} \cite{yefr8}.
 \sms

Consider the quantum particle with electric charge e, mass m, and generalized
momentum
 $$ P_k \equiv - i\hbar \frac{\partial }{{\partial x_k }} - \frac{e}{c}A_k $$
in the simplest quaternionic space (all the parameters are constant,
connection, non-metricity, torsion and  curvature equal to zero). Hamiltonian
of such particle in Q-metrics
 $$ H \equiv  - \frac{1}{{2m}}P_k P_m {\bf q}_k {\bf q}_m$$
is the exact copy of Hamilton function of Pauli equation
 $$ H = \frac{1}{{2m}}\left({\vec p - \frac{e}{c}\vec A} \right)^2
 - \frac{{e\hbar}}{{2mc}}\vec B \cdot \vec \sigma, $$
and the spin term "automatically"\, acquires a coefficient equal to Bohr
magneton.

 \ses
 5. {\it Young-Mills field strength.}
 \sms

If one constructs a "potential"\,  vector in an arbitrary quaternionic space
from Q-connection components $\Omega_{amn}$ (indices $a,b,c$ enumerate
coordinates of basic Q-space, indices $j,k,m,n$ enumerate vectors of tangent
triad)
 $$ A_{ka} \equiv \frac{1}{2}\varepsilon_{kmn} \Omega_{amn}, $$
and similarly construct a "field strength"\, vector
 $$ F_{kab}\equiv \frac{1}{2}\varepsilon _{kmn} r_{mnab}, $$
from quaternionic curvature components
 $$ r_{knab} = \partial_a \Omega_{bkn} - \partial_b \Omega_{akn}
 + \Omega_{ajn} \Omega_{bjk} - \Omega_{bjk} \Omega_{ajn}$$
then these two geometrical objects are interconnected in the similar manner as
the field strength and potential of the Young-Mills field
 $$ F_{kab} \equiv \partial_b A_{ka} - \partial_a A_{kb}
 + \varepsilon_{kmn} A_{ma} A_{nb}.$$(formula)
It should be stressed that for the Q-spaces with metric (not affine) connection
curvature (field strength) identically vanish.

 \esect{Discussion}

Quaternionic numbers of course are first of all mathematical objects, so the
problem of development of their algebra, analysis and geometry is
self-consistent. But history of modern science states that once the geometry,
in particular differential geometry, is discussed the presence of physics is
unavoidable. There is a known point of view that Einstein who suggested General
Relativity was a pioneer in geometrization of physics. But it is also known
that quite earlier Maxwell formulated his electrodynamics in terms of
quaternions convenient for description of 'etheric tensions' which were thought
to represent field strength vectors. But since that the geometrical language
has not been utilized for many decades.

The aspects of quaternionic mathematics given in this review once again draw
attention to 'genetic relations' between physics and geometry: from description
of frames rotations to quaternionic field structure phenomena in Pauli
equations and Young-Mills theory.

Wide variety of possibilities provided by Q-approach and derived within it
non-traditional physical models, like six-dimensional space-time or mentioned
above coincidences may lead to opinion that quaternions are still a
mathematical play, something like 'lego' elements, from which one can build
many exotic constructions.

As a comment there are the following two observations.

1. Producing non-standard physical models Q-method nonetheless allows to
successfully solve physical problems thus being a useful tool for practical
purposes. A typical example: inherited exponential character of representation
of simple rotations helps to simply formulate summation of different rotations,
including, of course, imaginary rotations, describing relativistic boosts.
Recall that in classical mechanics summation of ordinary rotations is quite a
task.

2. All physical quaternionic theories are not heuristically invented, but
appear naturally from fundamental mathematical lows, as though confirming
Pythagorean idea on "world -- number"\, dependence. Indeed, Q-algebra, the last
associative algebra, describes well physical quantities, all of them up to our
knowledge being associative with respect to multiplication: from observable
kinematic and dynamic one, to tensors and spinors incorporated in the theories.
All this gives a hope that further efforts in the research "quaternions --
physical laws"\, relations will once grow into wide scientific programme. Yet
another small, but persevering step in this direction has been recently made,
when the author of this review succeeded to found an exact solution for
relativistic oscillator problem in the framework Quaternionic Relativity.
Details of the solution will be published elsewhere.

 \end{document}